\documentclass[10pt,letterpaper]{article}
\usepackage[top=0.85in,left=2.75in,footskip=0.75in]{geometry}

\usepackage{pseudocode}
\usepackage{multirow}

\usepackage{amsmath,amssymb}

\usepackage{changepage}

\usepackage[utf8x]{inputenc}

\usepackage{textcomp,marvosym}

\usepackage{cite}

\usepackage{nameref,hyperref}

\usepackage[right]{lineno}

\usepackage{microtype}
\DisableLigatures[f]{encoding = *, family = * }

\usepackage[table]{xcolor}

\usepackage{array}

\newcolumntype{+}{!{\vrule width 2pt}}

\newlength\savedwidth

\newcommand\thickhline{\noalign{\global\savedwidth\arrayrulewidth\global\arrayrulewidth 2pt}%
\hline
\noalign{\global\arrayrulewidth\savedwidth}}

\raggedright
\usepackage[aboveskip=1pt,labelfont=bf,labelsep=period,justification=raggedright,singlelinecheck=off]{caption}

\makeatletter
\renewcommand{\@biblabel}[1]{\quad#1.}
\makeatother

\usepackage{lastpage,fancyhdr,graphicx}
\usepackage{epstopdf}
\fancyheadoffset[L]{2.25in}
\fancyfootoffset[L]{2.25in}
\begin{document}
\vspace*{0.2in}

\begin{flushleft}
{\Large
\textbf\newline{A statistical methodology for data-driven partitioning of infectious disease incidence into age-groups} 
}
\newline
\\
Rami Yaari\textsuperscript{1,2*},
Amit Huppert\textsuperscript{2,3},
Itai Dattner\textsuperscript{1}
\\
\bigskip
\textbf{1} Department of Statistics, University of Haifa, Haifa, 34988, Israel
\\
\textbf{2} Bio-statistical and Bio-mathematical Unit, The Gertner Institute for Epidemiology and Health Policy Research, Chaim Sheba Medical Center, Tel Hashomer, 52621, Israel
\\
\textbf{3} School of Public Health, the Sackler Faculty of Medicine, Tel-Aviv University, Tel Aviv, 69978, Israel
\\
\bigskip

* ramiyaari@gmail.com

\end{flushleft}
\section*{Abstract}
Understanding age-group dynamics of infectious diseases is a fundamental issue for both scientific study and policymaking. Age-structure epidemic models were developed in order to study and improve our understanding of these dynamics. By fitting the models to incidence data of real outbreaks one can infer estimates of key epidemiological parameters. However, estimation of the transmission in an age-structured populations requires first to define the age-groups of interest. Misspecification in representing the heterogeneity in the age-dependent transmission rates can potentially lead to biased estimation of parameters. We develop the first statistical, data-driven methodology for deciding on the best partition of incidence data into age-groups. The method employs a top-down hierarchical partitioning algorithm, with a metric distance built for maximizing mathematical identifiability of the transmission matrix, and a stopping criteria based on significance testing. The methodology is tested using simulations showing good statistical properties. The methodology is then applied to influenza incidence data of 14 seasons in order to extract the significant age-group clusters in each season.

\section*{Author summary}
The governing forces that shape the spread of infectious diseases are complex. Heterogeneity in the infection process due to age-dependent contact patterns is known to play a major role in this respect. In order to quantify the effect of age-dependent transmission and develop improved intervention policies for epidemics such as optimal vaccination programs, researchers employ mathematical models of disease spread, which parameterize the transmission among individuals according to their age-groups. One way to calibrate these models and estimate the age-dependent transmission parameters is by fitting the models to past incidence data of real outbreaks. Currently, however, there is no quantitative methodology for deciding how to partition incidence data into age-groups. Partitioning the data into too many age-groups would impede the proper estimation of the age-dependent transmission parameters. Partitioning the data into too few age-groups or not the ''correct'' age-groups, could give a distorted picture of the age-dependent infection dynamics. To fill the gap, we developed a statistical methodology that, given incidence data according to age, establishes the best partition of the data into age-groups in a way that would maximize our ability to estimate the transmission parameters from the data and obtain a proper picture of the age-dependent infection dynamics.


\section*{Introduction}\label{sec:inro}

The governing forces that shape infectious diseases spread are complex. Improving our knowledge of the relative importance of the different elements and their interplay are essential in order to obtain deeper understanding of the dynamics. Currently, the best tool to study such complex interactions are mathematical models.  Infectious disease modeling has a long history dating back to the seminal works of Daniel Brenoulli on variolation of smallpox~\cite{bernoulli1760essai}. From a theoretical perspective, models have revealed the existence of an outbreak threshold, which depends on features of the disease and the population~\cite{kermack1927contribution, kermack1932contributions, kermack1933contributions, anderson1992infectious}. Today it is realized that models are vital tools for identifying the underlying mechanisms giving rise to the observed dynamics. From a practical public health perspective, models are the backbone for developing better control strategies to aid in the mitigation and elimination of the disease, and to predict the unfolding in real-time of outbreaks. In order to obtain a 'useful' model one needs to balance between model simplicity, which enables to conduct a systematic mathematical analysis, with the essential biological complexity necessary for a realistic description of the infection process. For the purpose of parameter inference, the suitable level of model complexity strongly depends on the quality of the data. More so, the preprocessing of the data can potentially affect the appropriate level of model complexity.    

One example demonstrating the tension between constructing a simple or complex model stems from the empirical feature that in many infectious diseases, incidence is age-dependent. For instance, in~\cite{goldstein2015, goldstein2015b}
a method was developed for identifying the relative role of different age-groups in an epidemic from incomplete incidence data, finding children to have the higest relative risk in epidemics of influenza and pertussis.
Such observations has motivated the development of age-group models also known as age-stratified or age-structure models. These models are extensions of the classical deterministic SIR (susceptible - infected - removed) model~\cite{ross1911prevention, kermack1927contribution, kermack1932contributions, kermack1933contributions} - a common paradigm upon which more sophisticated models are based. Modifying the SIR model to account for age-dependent transmission requires employing an $M\times M$ transmission matrix ($M$ being the number of age-groups), known also as the WAIFW (Who Acquires Infection From Whom) matrix, describing the infection rates between the different groups~\cite{baguelin2011age, ott2012global, anderson1982control, schenzle1984age}. Ignoring the population age-structure can affect analysis results and potentially lead to biased estimation of parameters of interest. On the other hand, accounting for unnecessary age-groups (over stratifying the population) inflates the number of parameters which need to be inferred and can lead to both identification problems and a heavy computation burden. 

Parameter inference of mathematical epidemiological models is today a common practice~\cite{dattner2018modern} that requires fitting the models to real outbreak data~\cite{bolker2008ecological, morton2005discrete, ionides2006inference, hooker2010parameterizing}, a challenging task even in cases of low dimensional models~\cite{voit2004decoupling}. Using recent advances in the area of parameter estimation for differential equations in general~\cite{dattner2015optimal, dattner2017modelling}, and for SIR models in particular~\cite{dattner2018modern, yaarietal18}, it is possible to tackle the current problem of estimating parameters of a potentially large system of differential equations, which is a fundamental part of the problem of fitting age-group models. Specifically,~\cite{yaarietal18} demonstrated that it is possible to estimate the transmission matrix of two age-groups from their incidence data directly without strong assumptions regarding the structure of the matrix, as long as the incidence data of the two groups are sufficiently 'different' (see below). This raises the question, given incidence data, how can we partition the data into various age-groups in such a way that will allow us to estimate the transmission matrix accurately?

Currently, to the best of our knowledge, there is no data-driven criterion on how to partition incidence data into age-groups. Existing practices to define the age-groups used for analyzing or modeling incidence data include: i) fixed age-group blocks (e.g., groups of 5 years), ii) administrative divisions (e.g., infants, toddlers, school children, etc) and iii) division based upon age-dependent disease prevalence. The transmission matrix estimated using subjective or ad-hoc age-group partitioning can have an impact on both the perceived disease dynamics and on the estimates of key epidemiological parameters (e.g, the basic reproductive number $R_0$, defined as the mean number of infectives that a single case generates over the course of its infectious period in an entirely susceptible population). To overcome this gap we develop a statistical methodology for data-driven partitioning of infectious disease incidence into age-groups. The methodology is based on a criterion that is intended to maximize mathematical identifiability of the transmission matrix. Thus, the methodology can serve as a tool for the exploration and analysis of outbreak data that is useful for studying age-dependent transmission dynamics.

\section*{Materials and methods}
\subsection*{Problem definition}
As mentioned above, a classical model used to describe the spread of a single outbreak in a population is the susceptible-infected-recovered (SIR) model. An age-group SIR model depicts the epidemic spread in different age-groups. A formulation of the model for $j=1,...,M$ age-groups, using an ordinary differential equation (ODE) system, is given by:
\begin{equation}\label{eq:ode_sir_model}
\bigg\{
\begin{array}{l}
S_j^{\prime}(t)=-S_j(t)\sum_{k=1}^{M}\beta_{k,j}I_k(t)/N_k(t),
\\
I_j^{\prime}(t)=S_j(t)\sum_{k=1}^{M}\beta_{k,j}I_k(t)/N_k(t) -\gamma I_j(t),
\\
R_j^{\prime}(t)=\gamma I_j(t).
\end{array}
\end{equation}
Here, $\beta$ signifies the age-group transmission matrix, in which the element $\beta_{k,j}$ is the infection transmission rate for an infective individual of age-group $k$ and a susceptible individual of age-group $j$. The parameter $\gamma$ is the recovery rate, which is assumed here to be identical for all age-groups, and $N_j$ is the size of age-group $j$. The initial conditions for the system include the initial number of susceptibles  ${S_0}_j$ and the initial number of infected  ${I_0}_j$  in each age-group $j$ (the number of recovered at any time is given by $R_j(t)=N_j(t)-S_j(t)-I_j(t)$, since the model assumes a closed population). While $I_j(t)$ in this model represents the prevalence at time $t$, the incidence at discrete time periods $t'=[t-1,t]$ can be calculated using the model as $i_j(t')=S_j(t-1)-S_j(t)$. 

Typically, the incidence $i_j$ is observed with some noise. Assume that the observed incidence $\widetilde{i}_j$ is obtained from the actual incidence according to the following statistical model: 
\begin{equation}\label{eq:stat_model}
\widetilde{i}_j(t) =i_j(t) +\epsilon_j,\ \epsilon_j \sim N(0,\sigma_j^2),\  j=1,...,M,\  t=1,...,n.
\end{equation}
This simple statistical model is a good approximation for a Poisson distributed noise term for a large enough population.
However, other statistical models can also be considered and employed in a similar scheme. Given noisy incidence curves $\widetilde{i}_j(t)$ for $j=1,...,M$ age-groups, constituting the most detailed description available of the incidence data in different age-groups, our goal is to find a partition of the incidence data into $M_0\leq M$ age-groups, that still allows estimating the age-group SIR model parameters, and in particular, the age-group transmission matrix $\beta$. 

\subsection*{Partitioning algorithm}
Following is a description of the algorithm used to partition the incidence data into age-groups. The algorithm is based on a divisive (top-down) hierarchical clustering approach, where first a tree is built, representing a hierarchical partitioning of the data into the most basic age-groups, and then the tree is pruned using a significance testing scheme. Since both the building and the pruning of the tree are based on the same crtierion (given below), it is crucial that each of these steps (building and prunning) will be performed using a different data set generated from the same observed process. Multiple data sets for the same observed process can be obtained by dividing the incidence data into random sets (at the expanse of some loss in statistical power), or if there are multiple sources observing the same process (e.g., incidence data obtained from two different Health Maintainance Organizations). Another approach for obtaining multiple data sets is suggested in the application to influenza data section below. For now, let us assume that two data sets are given, so that we have noisy incidence data $\widetilde{i}_{j,k}(t)$ for $t=1,...,n$ time points, $j=1,...,M$ age-groups and $k=1,2$ sets, where it is assumed that $\widetilde{i}_{j,k}(t)=i_j(t) +\epsilon_{j,k}$ and $\epsilon_{j,k} \sim N(0,\sigma_j^2)$ for each $(j,k)$. One data set will be used to build the tree and the other one to prune it.

To build the tree the algorithm starts with the incidence data of all age-groups summed together into a single time-series. It then tests the $M-1$ possible ways to divide the incidence into two continuous age-group partitions, $i_{a,k}$ and $i_{b,k}$ (supressing the notation of $t$): $$P_k: \quad  i_{a,k}=\sum_{j=1}^k\widetilde{i}_j, \quad i_{b,k}=\sum_{j=k+1}^M\widetilde{i}_j.$$ 
For each candidate partition $P_k, k=1,...,M-1$, a statistic $q_k$ is calculated using $i_{a,k}$ and $i_{b,k}$. The statistic $q_k$ will be used for deciding whether a partition should be made at this $k$. To be more specific, in a previous work~\cite{yaarietal18}  found that having two age-group incidence curves that are the same up to a factor, leads to mathematical non-identifiability of the transmission matrix. When the two curves are almost similar (up to a factor), the transmission matrix, while mathematically identifiable, would be practically non-identifiable. The algorithm therefore makes use of this criterion as the basis for deciding how to separate the incidence data into age-groups. That is, a partition that gives rise to an identifiable model. For partition $P_k$, the null hypothesis $H_{0,k}$ is that the two underlying incidence curves are the same up to a factor, meaning that $i_{b,k}(\cdot)=ci_{a,k}(\cdot)$ for some constant $c$. We set $d_k(t)=i_{b,k}(t)-\hat{c}i_{a,k}(t)$,  where $\hat{c}=\sum_{t=1}^ni_{a,k}(t)i_{b,k}(t)/\sum_{t=1}^n(i_{a,k}(t))^2$. Given the statistical model (Eq~\ref{eq:stat_model}), $var(d_k(t))=\hat{c}^2\sigma_{a,k}^2+\sigma_{b,k}^2=:v_k$, where $\sigma_{a,k}^2=\sum_{j=1}^k\sigma_j^2$ and $\sigma_{b,k}^2=\sum_{j=k+1}^M\sigma_j^2$. The distance measure is then set as $q_k:= \frac{1}{v_k}\sum_{t=1}^n(d_k(t))^2$.
Under $H_{0,k}$, we have $d_k(t)\sim N(0,v_k)$ for all $t$ and therefore $q_k \sim \chi^2(n)$. Given partitions $P_k$, $k=1,...,M-1$, of  $i_{a,k}$ and $i_{b,k}$, the algorithm looks for $k$ that maximizes the distance between the two groups:
\begin{eqnarray}
\tilde{k}&:=&\arg\max_{k=1,...,M-1}q_k,
\end{eqnarray}
and the selected partition is given by $P_{\tilde{k}}$. This procedure then repeats recursively until the data is partitioned completely. Note that in case $\tilde{k}$ is not unique, one can arbitrarily choose between them. 

To prune the tree, the algorithm makes another pass on the tree starting from its root. At each node, the statistic $q_{\tilde{k}}$ is calculated for
the now given partition $P_{\tilde{k}}$ using the second data set. It uses the statistic $q_{\tilde{k}}$ in a $\chi^2$ test for the null hypothesis $H_{0,\tilde{k}}$. That is, for a given significance level $\alpha \in (0,1)$, it will reject $H_{0,\tilde{k}}$ if $p=1-F_{\chi^2}(q_{\tilde{k}},n)\leq \alpha$. If the null hypothesis is not rejected, all child nodes of this node are pruned, meaning that the incidence data at this node will be clustered together. Otherwise, the process continues with the child nodes. When the process stops, the leaves of the remaining pruned tree hold the clustering of the data.

Since multiple hypothesis tests are performed during the pruning of the tree, it is necessary to modify the significance level $\alpha$ accordingly. The algorithm employs the method described in~\cite{meinshausen2008hierarchical} to control the familywise error rate (FWER) at level $\alpha$ simultaneously across all nodes of the tree. At a given node that clusters $m$ groups of the original $M$ groups, the significance level according to~\cite{meinshausen2008hierarchical} should be set as: $\alpha^*=\frac{m}{M}\alpha$ (see also ~\cite{kimes2017statistical}). According to Theorem 1 of~\cite{meinshausen2008hierarchical} this modification will ensure that the probability for rejecting the null hypothesis at each node will be smaller than $\alpha$. It means that in the root node, the significance level is $\alpha$, but as we proceed down the tree and get to finer and finer resolutions, the significance level becomes smaller, making it harder to partition the smaller clusters. We note that in order for Theorem 1 of~\cite{meinshausen2008hierarchical} to hold in this case, we need monotonicity of p-values down the tree, which is guaranteed by the pruning procedure. 

So far it was assumed that the variance parameters $\sigma_1^2,...,\sigma_M^2$ are given.  In case they are unknown, they can be estimated from the $L=2$ data sets as $\hat{\sigma_j^2}=\frac{1}{n(L-1)}\sum_{t=1}^n\sum_{k=1}^L(\widetilde{i}_{j,k}(t)-\mu_j(t))^2$ where $\mu_j(t)=\frac{1}{L}\sum_{k=1}^L\widetilde{i}_{j,k}(t)$. Once the variance components are estimated, the tree can be built and pruned as described above. Note that if there are $L>2$ data sets, it is possible to estimate $\hat{\sigma_j^2}$ as above from all $L$ data sets, and then merge them into two data sets $L_1$ and $L_2$ ($L_1+L_2=L$) for building and prunning the tree, by setting $\widetilde{i}_{j,k}(t)=\frac{1}{L_k}\sum_{l=1}^{L_k}\widetilde{i}_{j,l}(t)$ and $\hat{\sigma_{j,k}^2}=\hat{\sigma_j^2}/{L_k}$ for $k=1,2$. In \nameref{S1_Appendix} another approach for estimating the variance parameters $\sigma_j^2$ and performing the partitioning algorithm with only a single data set is provided.

Algorithms~\ref{alg:build} and~\ref{alg:prune} provide a pseudo-code for the tree building and prunning procedures, using one data set ($X_1$) to build the tree and another one ($X_2$) to prune it. The columns of the matrices $X_1$ and $X_2$ are the observed incidence belonging to the current node, $\widetilde{i}_1,...,\widetilde{i}_m$. Input vector $V$ holds the noise parameters for the current node $\sigma_1^2,...,\sigma_{m}^2$. The output of the BuildTree procedure is the complete tree $T$ which is given as input to the PruneTree procedure, together with the number of groups $M$ and the significance level $\alpha$, in order to produce the pruned tree. Note that the two procedures can be performed in a single cycle, where at each node one data set is used to decide how to partition the data and the other one is used to decide whether or not to partition it.

\begin{pseudocode}[ruled]{BuildTree}{\text{$X_1,V$}}
\label{alg:build}
\text{$T=\{\}$} \\
\text{$n=nrow(X_1)$}, 
\text{$m=ncol(X_1)$}, \\ 
 \IF (\text{$m == 1$})  \THEN 
 	\text{$T.k=1$}; \text{return($T$)} \\
\text{$q = []$} \\
\FOR \text{$k$} \in \text{$1:(m-1)$} \DO \BEGIN 
	\text{$V_a=\sum_{i=1}^kV[i]$}, 
	\text{$V_b=\sum_{i=k+1}^{m}V[i]$} \\
	\text{$I_a=\sum_{i=1}^kX_1[,i]$}, 
	\text{$I_b=\sum_{i=k+1}^{m}X_1[,i]$} \\
	\text{$c=\sum_{t=1}^nI_a(t)I_b(t)/\sum_{t=1}^nI_a(t)^2$} \\
 	\text{$q[k]=\frac{1}{c^2V_a+V_b}\sum_{t=1}^n(I_b(t)-cI_a(t))^2$} \\
\END \\
\text{$\tilde{k} = argmax(q)$} \\
\text{$V_a=V[1:\tilde{k}]$}, 
\text{$V_b=V[(\tilde{k}+1):m]$}, \\
\text{$X_a=X_1[,1:\tilde{k}]$}, 
\text{$X_b=X_1[,(\tilde{k}+1):m]$}, \\
\text{$T.k=\tilde{k}$} \\
\text{$T.a=BuildTree(X_a,V_a)$} \\
\text{$T.b=BuildTree(X_b,V_b)$} \\
\text{return($T$)} \\
\end{pseudocode}

\begin{pseudocode}[ruled]{PruneTree}{\text{$T,X_2,V,M,\alpha$}}
\scriptsize
\label{alg:prune}
\text{$n=nrow(X_2)$}, 
\text{$m=ncol(X_2)$}, \\ 
 \IF (\text{$m == 1$})  \THEN 
 	\text{return($T$)} \\
\text{$k = T.k$} \\
\text{$V_a=\sum_{i=1}^kV[i]$}, 
\text{$V_b=\sum_{i=k+1}^{m}V[i]$} \\
\text{$I_a=\sum_{i=1}^kX_2[,i]$}, 
\text{$I_b=\sum_{i=k+1}^{m}X_2[,i]$} \\
\text{$c=\sum_{t=1}^nI_a(t)I_b(t)/\sum_{t=1}^nI_a(t)^2$} \\
\text{$q=\frac{1}{c^2V_a+V_b}\sum_{t=1}^n(I_b(t)-cI_a(t))^2$} \\
\text{$p = 1-F_{\chi^2}(q,n)$} \\
 \IF (\text{$p \leq \frac{m}{M}\alpha$}) \THEN \BEGIN
	\text{$X_a=X_2[,1:k]$}, 
	\text{$X_b=X_2[,(k+1):m]$}, \\
	\text{$V_a=V[1:k]$}, 
	\text{$V_b=V[(k+1):m]$}, \\
	\text{$T.a=PruneTree(T.a,X_a,V_a,M,\alpha)$} \\
	\text{$T.b=PruneTree(T.b,X_b,V_b,M,\alpha)$} \\
\END 
\ELSE 
	\text{$T=NULL$} \ \text{\# meaning remove split}\\
\text{return($T$)} \\
\end{pseudocode}

\subsection*{Partitioning algorithm with bagging }
Here we describe an enhancement to the partitioning algorithm that employs bootstrap aggregation (bagging). As before, it is assumed that there are noisy incidence data $\widetilde{i}_{j,k}(t)$ for $t=1,...,n$ time points, $j=1,...,M$ age-groups and $k=1,2$ sets. However, in this case, instead of running the algorithm once on the two data sets, the algorithm is run $B$ times, where each time it samples which incidences will be used for building the tree and which ones will be used for prunning it. That is, for each age-group $j$, it samples either $\widetilde{i}_{j,1}$ or $\widetilde{i}_{j,2}$ for the data set used for building the tree and the other one for the data set used for prunning the tree. In this way, a different set of data sets is generated for each run of the algorithm (in total there are $2\times2^M$ possible combinations for generating the two data sets and allocating them for building or prunning the tree). From these $B$ runs of the algorithm, $B$ clusterings are obtained, from which one is selected using the following procedure:
\begin{enumerate}
\item For each clustering, the adjusted Rand index (ARI)~\cite{rand1971}, measuring the similarity between two clusterings, is calculated for this clustering and each of the other $B-1$ clusterings.
\item The mean ARI (mARI) for each clustering is calculated as the mean of the $B-1$ ARI values calculated for it. 
\item The clustering with the highest mARI (i.e., the clustering with the best agreement to all other clusterings) is selected. 
\item If there are more than one clustering with the maximum mARI, then the one with least number of clusters (i.e., the more parsimonious clustering) is selected. 
\item If there are more than one clustering with the maximum mARI and the same minimum number of clusters, then the one that appeared the most within the $B$ clusterings is selected.
\end{enumerate}

\subsection*{Verifying Type-I error and power of the algorithm}
Monte-carlo simulations were used to test the Type-I error and power of the algorithm. The age-group SIR model (Eq~\ref{eq:ode_sir_model}) was ran to generate incidence data $i_1,...,i_M$ where $M=20$ or $M=40$ groups. A diagonal age-group transmission matrix $\beta$ was set to reflect various number of actual clusters $M_0$. For example, in order to generate incidence with a single cluster (i.e., $M_0=1$ clusters), all the diagonal components were set to the same value ($\lambda=0.84$) so that in actuality the same incidence is generated for all age-groups. To generate incidence with $M=20$ groups and $M_0=4$ clusters, the diagonal was set to
$[\lambda,\lambda,\lambda,\lambda,\lambda,(1-\delta)\cdot(\lambda,\lambda,\lambda,\lambda,\lambda),
\lambda,\lambda,\lambda,\lambda,\lambda,(1-\delta)\cdot(\lambda,\lambda,\lambda,\lambda,\lambda)]$ with $\delta>0$, 
so that in practice there are two different sets of incidence generated, separating the population into four distinct clusters (1-5,6-10,11-15,16-20).
In all simulations, the infection recovery rate was set to $\gamma=0.3$, the group sizes were set to $N=1$ and the initial conditions to $S_0=0.5$,  $I_0=5e-4$ for all $M$ groups. The effective reproductive number (the expected number of infectives infected by a single infective in a partially susceptible population) of group $j$ for which $\beta_{j,j}=\lambda$ is ${R_e}_j=\frac{\lambda}{\gamma}S_0=0.84/0.3\cdot 0.5=1.4$~\cite{keeling2008}. The effective reproductive number of group $k$ for which $\beta_{k,k}=(1-\delta)\cdot\lambda$ is ${R_e}_k={R_e}_j\cdot (1-\delta)$, implying that $\delta$ represents the percentage difference in $R_e$ between two age-group clusters. The number of observed time points in all simulations was set to $n=100$ and the significance level to $\alpha=0.05$.
In the Monte Carlo study, $1000$ sets of noisy incidence $\widetilde{i}_{1,k},...,\widetilde{i}_{M,k}$ with $k=1,...,L$ were generated from the incidence data $i_1,...,i_M$ using the observation model (Eq~\ref{eq:stat_model}) with the same $\sigma_j$ for all $j=1,...,M$.  The partitioning algorithm was run on each set of $1000$ simulations twice, once while assuming $\sigma_j^2$ are known and once while estimating $\sigma_j^2$ from the data (not assuming $\sigma_j^2$ are the same). When testing $\alpha$ we recorded the number of times the algorithm made a type-I error, \textit{i.e.}, it rejected $H_0$ and partitioned a cluster when it should not have. With $M_0=1$ this means that the algorithm would reject $H_0$ at the root node and split the incidence into more than a single cluster. With $M_0>1$ this means that the algorithm would reject $H_0$ at one of the nodes further down the tree representing an actual cluster. To test the power of the algorithm sets of $1000$ simulations were ran for varying values of $\delta$, and the number of times the algorithm found the correct clustering was recorded.  In addition, for each simulation, the ARI measuring the similarity between the actual clustering and the algorithm's output clustering was calculated, and the mean ARI (mARI) for each set of simulation was determined. The effect of employing bagging with various number of runs $B$ on the power of the algorithm was also tested, as well as the effect of increasing the number of available data sets ($L$).

The sensitivity of the partitioning algorithm to the assumption of a simple Gaussian noise was tested by generating observed age-group incidences using Poisson distributed noise term and examining its effect on the number of type-I errors and the power of the algorithm.These simulations were performed in the same manner described above except that, when running the transmission model, the age-group sizes were set to $N_j=10^4$ for $j=1,..,M$ instead of $N_j=1$, and an observation model $\widetilde{i}_j(t)=Poisson(i_j(t))$ was employed in order to generate the observed incidence instead of model (Eq~\ref{eq:stat_model}).

\subsection*{Application to influenza incidence data}
The clustering methodology was applied to weekly incidence of influenza like-illness (ILI) diagnoses during 14 seasons (1998-2011). The diagnoses were given by physicians of the Maccabi Health Maintenance Organization in Israel, which serves approximately 25\% of the the Israeli population. The time period for each season was defined according to results of virological tests conducted on samples taken from patients visiting sentinel clinics. For the 2009 pandemic season,  only the period of the winter wave starting from October 2009 was used. The time period used per season varied between 11 and 18 weeks (mean=14.6, s.d.=1.8).  In order to obtain multiple incidence data sets per age-group, we considered basic age-groups of 2-year bands, 3-year bands and 4-year bands. With 2-year bands the basic age-groups were 0-1,2-3,4-5,...,64-65+ for a total of $M=33$ groups, with 3-year bands the basic age-groups were 0-2,3-5,6-8,...,63-65+  for a total of $M=22$ groups, and with 4-year bands the basic age-groups were 0-3,4-7,8-11,...,60-64+  for a total of $M=16$ groups. In each case, the assumption was that the incidences within a basic age-group (e.g., the incidences for age 0 and age 1 within the basic age-group 0-1 in the case of 2-year bands) are the same up to some noise (due to measurement error) and possibly scaling (due to differences in reporting rates). To address differences in scaling, the incidence curves within a basic age-group with $L$ incidences were scaled by setting $\check{i}_j(\tau)=\widetilde{i}_j(\tau)\frac{\sum_{k=1}^{L}\sum_{t=1}^{n}\widetilde{i}_k(t)}{L\times\sum_{t=1}^{n}\widetilde{i}_j(t)}$, for $j=1...L$, $\tau=1,...,n$. For each season and each value of $L$, the partitioning algorithm was run $B=1000$ times, where for each run of the algorithm, two of the $L$ incidences per basic age-group were randomly selected and were used to estimate the variance within the basic age-groups. Then one set of incidences was allocated for building the hierarchical clustering tree and the other one for pruning it. Finally, the procedure described in the subsection above ('Partitioning algorithm with bagging') was used to select the final partition per season.

\section*{Results}
\label{sec:results}
\subsection*{Simulations}
Table~\ref{tab:alpha} summarize the results of the simulations to verify type-I error. As can be seen, when $\sigma_j^2$ are known, type-I error occures less than $\alpha=0.05$ of the time, as expected. When $\sigma_j^2$ are unknown and are estimated, the increase in the occurences of type-I error are small. Table~\ref{tab:power} summarizes the results of the power simulations. With a large enough $\delta$ (the percentage difference in $R_e$ between two age-group clusters) relative to the selected intermediate value of $\sigma_j^2$, the power of the algorithm is high. With $\delta$ decreasing, the power decreases slowly until it falls off rapidly at some point, depending on the exact scenario. Estimating $\sigma_j^2$ does not have a significant effect on the power.  

\begin{table}[!ht]
\centering
\caption{
{\bf Results of simulations to verify type-I error.}
$M$ is the total number of age-groups. $M_0$ is the number of clusters. $\sigma_j^2$ is the Gaussian variance for age-group $j$, set to be the same for each group in these simulations (but are not assumed to be the same when estimated). The number of observed incidences per group was set to $L=2$ and the significance level set to $\alpha=0.05$.}
\label{tab:alpha}
\begin{tabular}{lcc|ccc}
\hline
\multicolumn{3}{c}{}&
\multicolumn{1}{c}{$\sigma_j^2$ known}&
\multicolumn{1}{c}{$\sigma_j^2$ unknown}\\	
\hline
\multicolumn{1}{c}{$M$}&	
\multicolumn{1}{c}{$M_0$}&	
\multicolumn{1}{c}{$\sigma_j^2$}&	
\multicolumn{2}{c}{type-I error} \\
\thickhline	
\multirow{6}{*}{20} &				
1
     &$10^{-6}$ & 0.033 &  0.054  \\ 
  & &$10^{-5}$ & 0.034 &  0.054  \\ 
\cline{2-5}
&2
     &$10^{-6}$ & 0.037 &  0.053 \\  
  & &$10^{-5}$ & 0.037 &  0.057  \\  
\cline{2-5}
&4
     &$10^{-6}$ & 0.035 &  0.046  \\ 
  & &$10^{-5}$ & 0.037 &  0.048  \\ 
\hline
\multirow{6}{*}{40} &				
1
     &$10^{-6}$ & 0.043 &  0.052  \\ 
  & &$10^{-5}$ & 0.043 &  0.051  \\ 
\cline{2-5}
&2
     &$10^{-6}$ & 0.038 &  0.066 \\  
  & &$10^{-5}$ & 0.040 &  0.063 \\  
\cline{2-5}
&4
     &$10^{-6}$ & 0.047 &  0.060 \\  
  & &$10^{-5}$ & 0.051 &  0.062  \\  
\end{tabular}
\end{table}

\begin{table}[!ht]
\centering
\caption{{\bf Results of power simulations.} $M$ is the total number of age-groups. $M_0$ is the number of clusters. $\delta$ is the distance between clusters, measured as percentage difference in $R_e$. Here, the noise parameters were set to $\sigma_j^2=5\times10^{-6}$ and the number of observed incidences per group was set to $L=2$.}
\label{tab:power}
\begin{tabular}{lcc|cc|ccc}
\multicolumn{3}{c}{}&
\multicolumn{2}{c}{$\sigma_j^2$ known}&
\multicolumn{2}{c}{$\sigma_j^2$ unknown}\\	
\hline
\multicolumn{1}{c}{$M$}&	
\multicolumn{1}{c}{$M_0$}&	
\multicolumn{1}{c}{$\delta$}&	
\multicolumn{1}{c}{power}&
\multicolumn{1}{c}{mARI} &
\multicolumn{1}{c}{power}& 
\multicolumn{1}{c}{mARI} \\
\thickhline	
\multirow{15}{*}{20} 
     & &$0.01$ & 0.248 & 0.422 & 0.242 &  0.427  \\ 
     & &$0.02$ & 0.901 & 0.981 & 0.879 &  0.977  \\ 
     &2&$0.03$ & 0.961 & 0.993 & 0.941 &  0.990  \\ 
     & &$0.04$ & 0.962 & 0.993 & 0.945 &  0.991  \\ 
     & &$0.05$ & 0.962 & 0.993 & 0.945 &  0.991  \\ 
\cline{2-7}
     & &$0.01$ & 0 & 0.028 & 0 &  0.023  \\ 
     & &$0.02$ & 0.091 & 0.329 & 0.088 &  0.326  \\ 
     &4&$0.03$ & 0.642 & 0.837 & 0.622 &  0.834  \\ 
     & &$0.04$ & 0.896 & 0.981 & 0.889 &  0.982  \\ 
     & &$0.05$ & 0.955 & 0.996 & 0.941 &  0.995  \\ 
\cline{2-7}
     & &$0.02$ & 0 & 0.014 & 0 &  0.016  \\ 
     & &$0.04$ & 0.124 & 0.363 & 0.116 &  0.360  \\ 
     &8&$0.06$ & 0.706 & 0.881 & 0.624 &  0.851  \\
     & &$0.08$ & 0.924 & 0.986 & 0.889 &  0.981  \\
     & &$0.10$ & 0.952 & 0.995 & 0.928 &  0.993  \\ 
\hline
\multirow{15}{*}{40} 
     & &$0.01$ & 0.503 & 0.860 & 0.491 &  0.858  \\ 
     & &$0.02$ & 0.928 & 0.990 & 0.904 &  0.989  \\ 
     &2&$0.03$ & 0.958 & 0.994 & 0.935 &  0.992  \\ 
     & &$0.04$ & 0.958 & 0.994 & 0.934 &  0.992  \\ 
     & &$0.05$ & 0.958 & 0.994 & 0.936 &  0.992  \\ 
\cline{2-7}
     & &$0.01$ & 0 & 0.080 & 0 &  0.076  \\ 
     & &$0.02$ & 0.404 & 0.763 & 0.389 &  0.760  \\ 
     &4&$0.03$ & 0.850 & 0.990 & 0.826 &  0.988  \\ 
     & &$0.04$ & 0.928 & 0.996 & 0.912 &  0.996  \\ 
     & &$0.05$ & 0.948 & 0.997 & 0.928 &  0.996  \\ 
\cline{2-7}
     & &$0.02$ & 0.004 & 0.088 & 0.001 &  0.088  \\ 
     & &$0.04$ & 0.803 & 0.985 & 0.762 &  0.982  \\ 
     & 8 &$0.06$ & 0.947 & 0.998 & 0.936 &  0.998  \\
     & &$0.08$ & 0.952 & 0.998 & 0.943 &  0.998  \\
     & &$0.10$ & 0.953 & 0.998 & 0.944 &  0.998  \\ 
\end{tabular}   
\end{table}

\begin{figure}[!h]
\caption{{\bf Effect of bagging and the number of observed incidence sets per age-group on the power of the partitioning algorithm.}
Panel 'A' shows the effect of bagging with number of bootstrap runs $B$. Panel 'B' shows the effect of the number of observed incidences per age-group $L$. The parameters used in these simulations were $M=20$, $M_0=8$ and $\sigma_j^2=5\times10^{-6}$ ($\sigma_j^2$ are assumed to be unknown). When testing the effect of bagging, $L$ was set to $2$. When testing the effect of the number of observed incidence curves per age-group, $B$ was set to $1$ (not using bagging).}
\label{fig:power}
 \begin{center}
 \includegraphics[width=120mm,height=60mm]{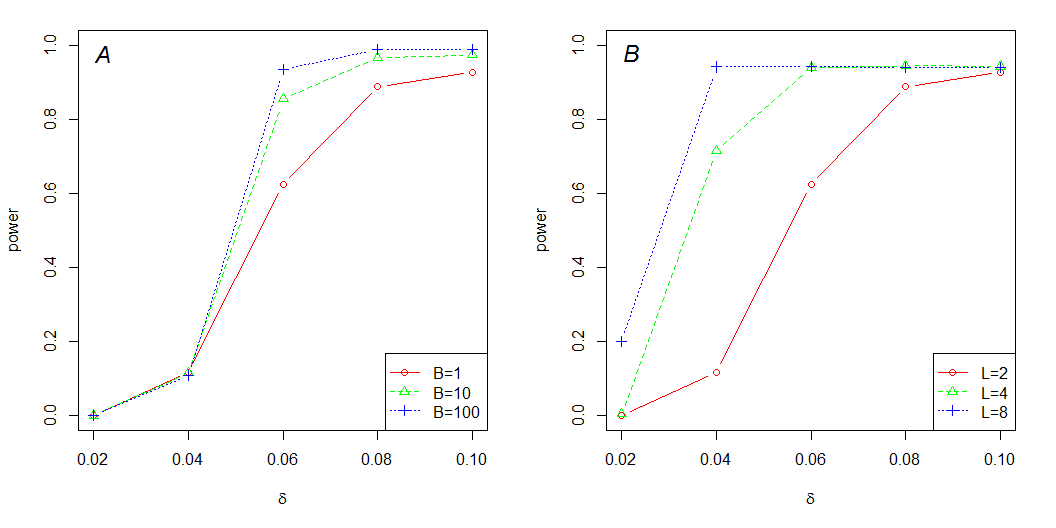}
 \end{center}
\end{figure}

Fig~\ref{fig:power} shows the effects of employing bagging (panel 'A') and using a larger number of observed incidence curves per age-group (panel 'B') on the power of the algorithm. Employing bagging does not seem to increase the power for small values of $\delta$, when the signal-to-noise ratio is low and the power of the algorithm without bagging is very low. It does however increase the power of the algorithm for larger values of $\delta$, when the power is intermediate or even high. It does so mainly by reducing type-I errors, that is, avoiding separating into different clusters incidences that should actually be grouped together. On the other hand, having a larger number of incidence curves per age-group increases the power of the algorithm significantly for smaller values of  $\delta$, but does not improve the power beyond its peak, when the power of the algorithm is already high.

As for the sensitivity of the algorithm to the assumption of a simple Gaussian noise, our results indicate that the basic algorithm still has good power in this case, whereas the occurance of type-I errors are not kept to below $\alpha$. However, using bagging with just $B=10$ runs, overcomes the problem with the type-I errors while maintaining high power for the algorithm (see \nameref{S3_Table} and \nameref{S4_Table}).

\subsection*{Application: Seasonal Influenza}

\begin{figure}[!h]
\caption{{\bf Clustering tree for the age-group incidence of the 1998 influenza season.} On the left, the tree obtained before prunning. On the right, the tree after prunning. Inner nodes are colored in orange and leaves in green. The leaves of the pruned tree contain the final clustering. }
\label{fig:clust_trees_1998}
 \begin{center}
 \includegraphics[width=140mm,height=84mm]{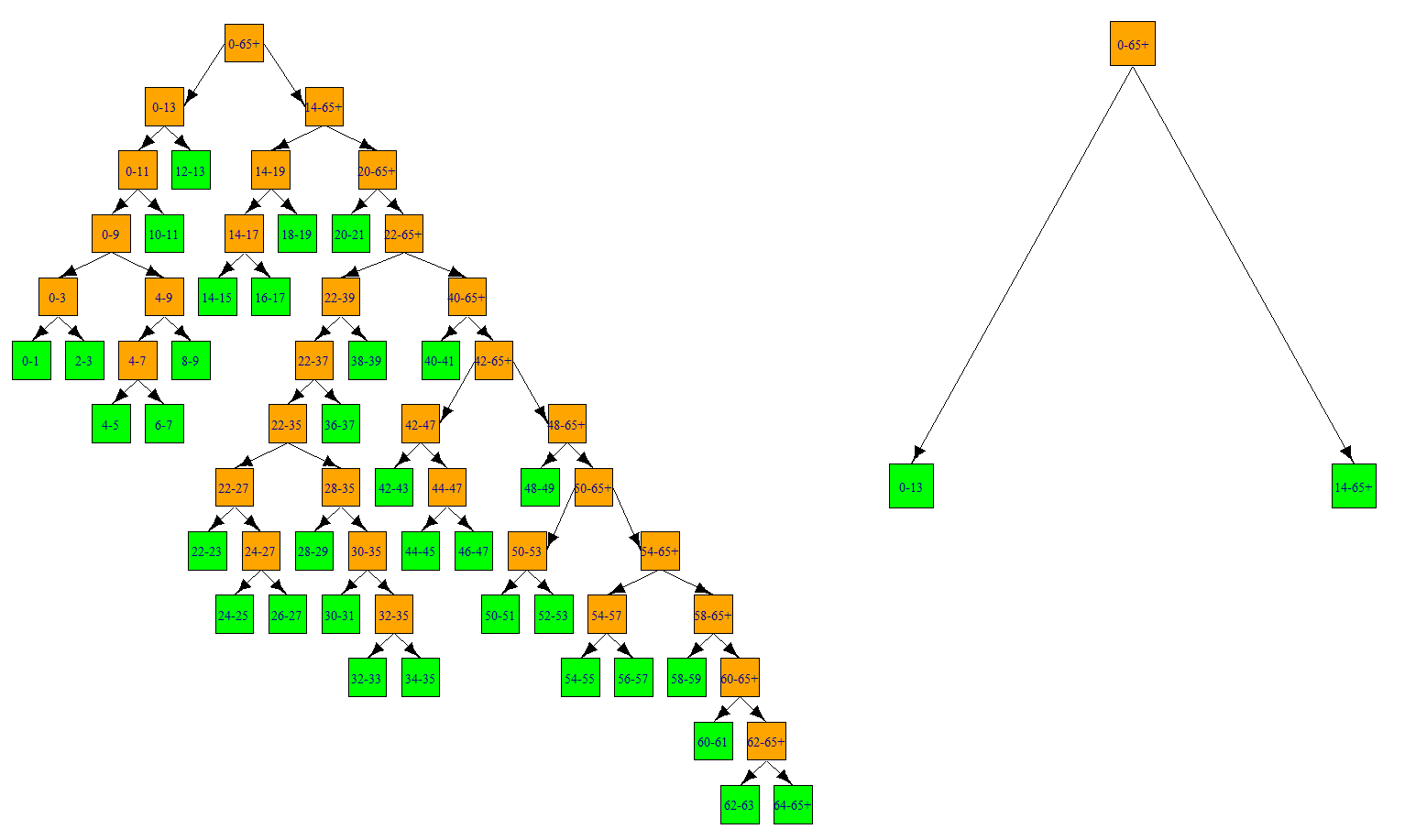}
 \end{center}
\end{figure}

Fig~\ref{fig:clust_trees_1998} presents the clustering tree before and after pruning, obtained by the partitioning algorithm by one of the 1000 boostrap runs on the weekly age-group ILI incidences of the 1998 influenza season, starting from basic age-groups of 2-year bands (0-1,2-3,4-5,...). The algorithm divided the incidence data into just two clusters: 0-14 and 15+. In fact, in this season, 398 out of the 1000 runs resulted in this clustering. The mARI for this clustering, measuring its agreement with all the other obtained clusterings, was 0.95, the highest of all the clusterings in this case, leading to the selection of this partition by the algorithm.

\begin{figure}[!h]
\caption{{\bf Age-group clusters obtained for each influenza season.} A visualization of the results presented in Table~\ref{tab:clusters}, starting from basic age-groups of 2-year bands. Red colors indicate clusters in children while blue colors indicate clusters in adults.}
\label{fig:clusters}
 \begin{center}
 \includegraphics[width=115mm,height=74mm]{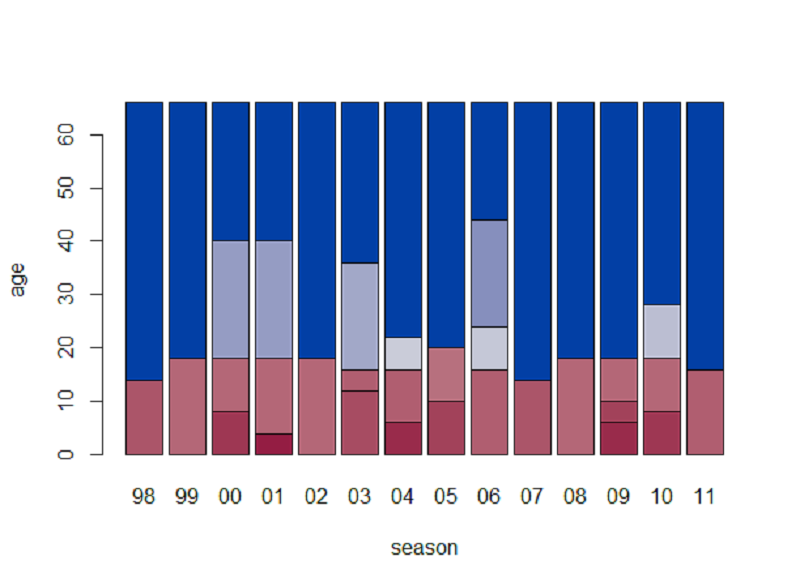}
 \end{center}
\end{figure}

\begin{figure}[!h]
\caption{{\bf Clustered influenza age-group incidences obtained for the 1998 influenza season.} The partitioning algorithm was ran starting from basic age-groups of 2-year bands. The incidences were normalized (each one divided by its sum) in order to present them in the same scale. In panel 'A' all age-group incidences are plotted together in one color, panel 'B' plots the incidences belonging to different clusters in different colors, while panel 'C' plots the mean incidence for each age-group cluster.}
\label{fig:clust_inc_1998}
 \begin{center}
 \includegraphics[width=70mm,height=140mm]{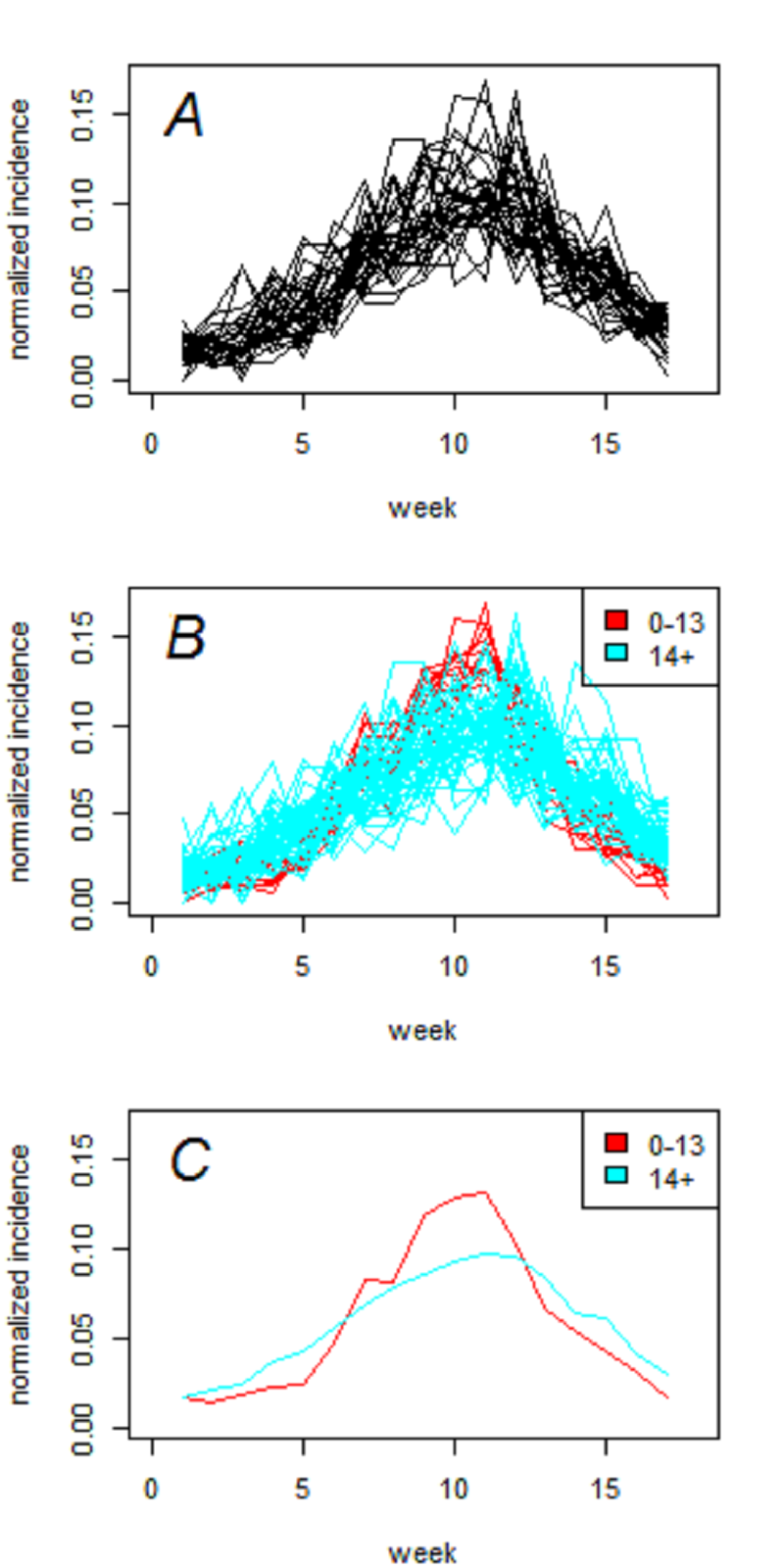}
 \end{center}
\end{figure}

Table~\ref{tab:clusters} present the clustering for each of the 14 influenza seasons in our data set, starting from basic age-groups of 2-year bands, 3-year bands and 4-year bands, obtained as described above for the 1998 season. The table also includes the mARI value in each case, which can serve as a measure of confidence in the results of the clustering. The number of clusters obtained for each season ranges between two and four. Using 2-year bands we tend to obtain more clusters as there are more groups to begin with ($M=33$). Nevertheless, in general, the results obtained using 2-year bands, 3-year bands and 4-year bands are very similar, demonstrating the results are robust to this pre-processing choice. Fig~\ref{fig:clusters} shows a visualization of the clustering obtained using 2-year bands. In all seasons, there is a partition of the incidences to children and adults, somewhere between age 12 and age 20. This is typically the most significant partition obtained by the clustering algorithm (the first partition of the clustering tree). In addition, in half of the seasons, the children are partitioned further to younger and older children, and in a third of the seasons, the adults are partioned further, mostly around the age of 40. Fig~\ref{fig:clust_inc_1998} presents the clustering obtained using the 2-year bands for the age-group incidences of the 1998 influenza season (see \nameref{S1_Fig} for similar outputs for all seasons).

\begin{table}[!ht]
\centering
\caption{{\bf Age-group clusters obtained for each influenza season}. The partitioning algorithm was ran with bagging, starting from basic age-groups of 2-year bands (0-1,2-3,4-5,...), 3-year bands (0-2,3-5,6-8,...) and 4-year bands (0-3,4-7,8-11,...). The clustering given here were the ones selected from the $B=1000$ runs according to the selection procedure detailed in the Methods section. The clusters are given by their separating ages (e.g., '14' means two clusters: 0-13 and 14+). The mean ARI (mARI) indicates the agreement of the clustering with the other clusterings obtained from the bootstrap runs.  }
\label{tab:clusters} 
\begin{tabular}{l|cc|cc|cc|c}
\multicolumn{1}{c}{}&
\multicolumn{2}{c}{2-year bands ($M=33$)}&
\multicolumn{2}{c}{3-year bands ($M=22$)}&
\multicolumn{2}{c}{4-year bands ($M=16$)}\\	
\hline
\multicolumn{1}{c}{season}&	
\multicolumn{1}{c}{clusters}&	
\multicolumn{1}{c}{mARI}&
\multicolumn{1}{c}{clusters}&	
\multicolumn{1}{c}{mARI}&
\multicolumn{1}{c}{clusters}&	
\multicolumn{1}{c}{mARI}\\
\thickhline	
1998 &  14 & 0.95 & 15  & 0.98  & 16 & 0.95 \\                          
1999 &  18 & 0.88 &  18 & 0.88  & 16 &  0.80\\
2000 &  8, 18, 40 & 0.91 & 12, 39  & 0.79 & 8, 40 & 0.75 \\ 
2001 &  4, 18, 40 & 0.86 & 15, 42  &  0.79 & 16 &  0.74 \\ 
2002 &  18 & 0.97 &  18  &  0.95 & 16 & 0.95 \\
2003 &  12, 16, 36 & 0.88 & 12, 18  & 0.83 & 12, 20 &  0.84 \\
2004 & 6, 16, 22 & 0.94 & 6, 18  & 0.88 & 16 &  0.91 \\
2005 & 10, 20 & 0.95 &  21 & 0.95 & 20 & 0.94 \\
2006 & 16, 24, 44 & 0.88 & 15, 27  & 0.87 & 16, 24 &  0.85 \\
2007 & 14 & 0.96 & 12  & 0.93  & 12 & 0.92 \\
2008 & 18 & 0.93 & 18  & 0.92 & 20 & 0.89 \\
2009 & 6, 10, 18 & 0.93 & 6, 18  & 0.86 & 8, 20 & 0.84 \\
2010 & 8, 18, 28 & 0.89 & 18, 30 & 0.85 & 12, 28 & 0.86 \\
2011 & 16 & 0.81 & 15 & 0.81 & 16 & 0.83 \\
\end{tabular}   
\end{table}	

\begin{figure}[!h]
\caption{{\bf Estimates of $R_0$ and $R_e$ for each influenza season.} The estimates were obtained by fitting the age-group SIR model to the clustered age-group incidence (black 'o's). These estimates are compared to estimates obtained by fitting a homogeneous transmission model to the incidence in the population as a whole (red 'x's).}
\label{fig:estimates_comparison}
 \begin{center}
 \includegraphics[width=100mm,height=100mm]{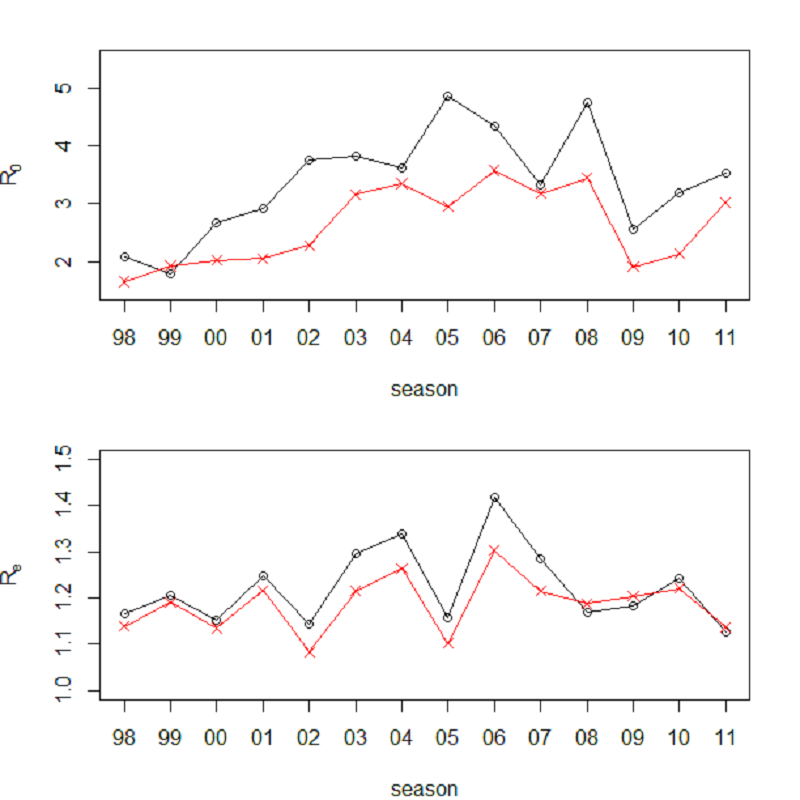}
 \end{center}
\end{figure}

In order to test if the transmission matrix $\beta$ can be estimated from the clustered age-group incidences, the age-group SIR model (\ref{eq:ode_sir_model}) was fitted to the clustered influenza incidence data of each season. The fitting was performed using the R-package simode~\cite{yaari2018textbf}. More details on the fitting are given in \nameref{S1_Appendix}. \nameref{S2_Fig} presents the obtained fits while the obtained parameters estimates are given in \nameref{S5_Table}. Using the estimated parameters we calculated the basic reproductive number $R_0=\frac{1}{\gamma}\rho(\mathcal{M}_0)$ and the effective reproductive number $R_e=\frac{1}{\gamma}\rho(\mathcal{M}_e)$ for each season. Here, $\mathcal{M}_0$ and $\mathcal{M}_e$ are the matrices whose entries are $\beta_{ij}\frac{N_i}{N_j}$, and $S_{0_i}\beta_{ij}\frac{N_i}{N_j}$, respectively, while $\rho$ denotes the spectral radius or maximum eigen value of the matrix~\cite{diekmann1990}. The estimates of $R_0$ and $R_e$ are given in Fig~\ref{fig:estimates_comparison}. In addition, for comparison, the estimates of $R_0$ and $R_e$ obtained by fitting an SIR model without age-groups to the observed incidence in the population as a whole were also plotted. As can be seen in the figure, the estimates of $R_0$ and $R_e$ obtained from the model fits to the age-group incidence are typically higher than the estimates obtained from the fits to the incidence of the population as a whole (see also~\cite[p.~61]{keeling2008}). 

\section*{Discussion}
\label{sec:disc}

In this paper we develop a statistical methodology for data-driven partitioning of infectious disease incidence into age-groups. To the best of our knowledge this is a first attempt at this task. The methodology is based on explicit mathematical models describing infectious disease dynamics~\cite{anderson1992infectious, keeling2011modeling, grassly2008mathematical, huppert2013mathematical, heesterbeek2015modeling}. It uses a criterion intended to enable parameter estimation of the transmission matrix from the incidence data, hence leading to identifiable models. 

Mathematical models have proven to be an effective tool for examining and exploring the dynamics of the spread of infectious diseases~\cite{ferguson2001foot, keeling2003modelling, lipsitch2003transmission, earn2012effects, yaari2016modeling}. In recent years, models have been applied in real-time as a supporting tool for decision makers to study and explore possible control and mitigation strategies~\cite{ferguson2001foot, keeling2003modelling, lipsitch2003transmission, earn2012effects, yaari2016modeling}. However, previous age-dependent modeling studies did not use a statistical methodology to stratify the population into age-groups. Using 'non-optimal' partitioning can lead to biases in critical parameter estimations. For instance, Fig~\ref{fig:estimates_comparison} demonstrates that ignoring age-group dynamics altogether can lead to reduced estimates of $R_0$ and $R_e$. This would translate to lower estimates of the herd immunity required to mitigate an outbreak, which can lead to non-optimal policy decisions. On the other hand, employing too many age-groups unnecessarily, can lead to difficulties in parameter inference (due to a larger number of model parameters), identifiability issues, as well as computational difficulties.         

The statistical methodology developed in this paper is based on significance testing in clustering, which addresses an important aspect of cluster validation~\cite{hennig2015handbook}. As mentioned in the aforementioned book, many cluster analysis methods will deliver clusterings even for homogeneous data. They assume implicitly that a clustering has to be found, regardless of whether this is meaningful or not. Indeed, in our case one would like to know how to distinguish between a clustering that reflects meaningful heterogeneity in the data and not just an artificial clustering of homogeneous data. Significance tests are the standard statistical tools for such distinctions and therefore we adopt this approach. However, in view of the complexity of clustering problems, we prefer to consider the methodology developed here as a tool for data exploration with $p$-values used as a threshold supporting scientific reasoning (see, e.g.,~\cite{pv}). As such, the methodology allows to conduct a systematic data exploration of the partitioning.

The methodology is a top-down hierarchical partitioning algorithm with a constraint on the type of divisions allowed and a statistically-based stopping criteria. The algorithm is designed to consider in a group only consecutive ages, hence enforcing a constraint on the partitioning. Thus, it is a semi-supervised hierarchical clustering (i.e., clustering with knowledge-based constraints, see. e.g.~\cite{zheng2011semi, bade2008creating,  zhao2010hierarchical}). Given the SIR model one could think of a model-based clustering~\cite{hennig2015handbook}. Unfortunately, model based clustering in such a case requires to calculate maximum-likelihood of the data and therefore one would need to solve numerically the differential equations for all potential partitions, which is numerically and computationally challenging. Thus the methodology only uses the identifiability criterion for partitioning, which seems to capture the main characteristics of the underlying dynamics and lead to good statistical properties such as power and ARI. 

The incorporation of bagging helps in reducing type-I errors. When applying the method to influenza data, bagging helped to stabilize the variance in the results obtained by running the algorithm using different pre-processing options (i.e., the choice of the starting basic age-groups), as well as the variance in the results obtained for different seasons. Our results indicate that the most significant partition of the incidence data is the partition to children (up to ages 14-18) and adults. Still, as can be seen in Fig~\ref{fig:clusters}, there is some variance in the clustering results across different seasons. A reasonable next step would be to try and explain this variance. It would be of epidemiological importance to test if the variability in age-group partitioning is an inherent property of influenza (e.g., due to the rapid evolution of the virus). Another explanation would be that we observe practical identifiability issues, meaning that several models are plausible given the data. Also, one may suggest to consider variability of $S_0$ within groups (we assume they are the same). Such questions require further research. Finally, an additional interesting future research direction would be to apply the algorithm developed here for clustering spatial incidence. This might require modifications in the distance metric used here as well as other considerations which are beyond the scope of this work.

\break

\section*{Supporting information}


\paragraph*{Appendix S1}
\label{S1_Appendix}
{\bf Supplementary information.} 

\paragraph*{Fig S1}
\label{S1_Fig}
{\bf Clustered influenza incidence for each influenza season.} 

\paragraph*{Fig S2}
\label{S2_Fig}
{\bf Model fits to clustered influenza incidence of each season.} 

\paragraph*{Table S1}
\label{S1_Table}
{\bf Results of simulations to verify type-I error using alternative methodology for a single data set.} 

\paragraph*{Table S2}
\label{S2_Table}
{\bf Results of power simulations using alternative methodology for a single data set.} 

\paragraph*{Table S3}
\label{S3_Table}
{\bf Results of simulations to verify type-I error using a Poisson distributed noise term.} 

\paragraph*{Table S4}
\label{S4_Table}
{\bf Results of power simulations using a Poisson distributed noise term.} 

\paragraph*{Table S5}
\label{S5_Table}
{\bf Estimated age-group SIR model parameters for each influenza season.}



\end{document}